\documentclass{ws-ijmpe}
\usepackage[super,compress]{cite}
\usepackage{amsmath, amssymb}
\usepackage{braket}
\begin{document}



\title{Cold Atom Quantum Simulator for Dilute Neutron Matter}

\author{Munekazu Horikoshi}

\address{Institute for Photon Science and Technology, Graduate School of Science,\\
The University of Tokyo\\
7-3-1 Hongo, Bunkyo-ku, Tokyo 113-8656, Japan\\
hori@psc.t.u-tokyo.ac.jp}

\author{Makoto Kuwata-Gonokami}

\address{Department of Physics, Graduate School of Science,\\
The University of Tokyo\\
7-3-1 Hongo, Bunkyo-ku, Tokyo 113-8656, Japan}

\maketitle


\begin{abstract}
The internal structure of neutron stars and the physical properties of nuclei depend on the equation of state (EOS) of neutron matter. Dilute neutron matter is a quantum system of spin-1/2 Fermi particles interacting via s-wave scattering. Although a nuclear system and an ultracold atomic system have length scales and energy scales that differ by several orders of magnitude, both systems follow a common universal EOS considering their non-dimensional universal interaction parameters. In this study, we determine the EOS of neutron matter in the dilute region, where the influence of the s-wave scattering length is dominant but that of the effective range is small, by utilizing a quantum simulator of ultracold $^6$Li atoms with Feshbach resonance.
\end{abstract}




\section{Introduction: neutron matter and neutron stars \label{Sec1}}

A neutron is an electrically neutral spin-1/2 Fermi particle with a mass of 940~MeV. Neutrons interact with the short-range interaction potential of the nuclear force with $r_0\sim2$~fm interaction range \cite{1}. Although it is difficult to determine the nuclear force strictly as a function of the distance between neutrons, the s-wave scattering length and the effective range have been determined to be $a_s=-18.9(4)$~fm \cite{2} and $r_e =2.75(11)$~fm \cite{3}, respectively. If neutrons were trapped in a potential and cooled below the Fermi temperature, what kind of density-dependent physical properties would they exhibit? And what would change if neutrons had different scattering parameters? Answering these questions about neutrons would help clarify the internal structure of neutron stars and the physical properties of nuclei.

Neutron stars have been confirmed by the detection of neutron star pulsars \cite{4}, X-ray astronomy observations \cite{5}, and gravitational wave observations \cite{6}. The average mass of observed neutron stars is approximately 1.4 times the solar mass \cite{7}. Neutron stars are considered to have radiuses of about 10 km, so neutron stars contain extremely dense matter, greater than the nuclear saturation density of $n_0=0.16$~fm$^{- 3}$. (In this paper, the number density is expressed as $n$ and the mass density is expressed as $\rho =mn$ where $m$ is the mass.) The temperature of a neutron star is extremely high at $T=10^5$--$10^7$~K \cite{8}, but the particle systems correspond to the zero-temperature limit for Fermi particles because the temperature range is several orders of magnitude lower than the Fermi temperature $T_F$ determined by the density. Therefore, neutron stars can be regarded as huge quantum stars that exhibit a remarkable quantum effect. Inside neutron stars, the state of matter continuously changes from a dilute region ($n \ll n_0$) to a dense region ($n>n_0$). Thus, neutron stars consist of hadron matter, unlike terrestrial matter consisting of atoms and molecules. The relation between density and pressure of particles $P(n)$, or the relationship between density and internal energy $E(n)$, is called the equation of state (EOS). To calculate the quantum state (structure) and eigenvalue (internal energy) at a given density for a hadron system, it is necessary to solve the quantum many-body problem considering various hadron interactions. Therefore, determining the EOS for neutron stars is a challenging subject in hadron physics.

The EOS can be transformed into the relationship between the radius and the mass of the star, called the M-R curve, by using the Tolman--Oppenheimer--Volkoff equation \cite{47,48}. According to this principle, there are as many M-R curves as there are EOSes \cite{7}. Thus, the EOS whose M-R curve passes through all the observed data of a neutron star represents the structure inside the neutron star most appropriately. Currently the measured data on neutron star radii is limited \cite{9}, though the amount of data will increase with future astronomical observations. Additionally, future measurements of the waveforms of the gravitational waves generated by neutron star binaries will help reveal the inner structure of neutron stars \cite{10}. 

The EOS of neutron matter gives the basic physical properties of the inner cluster and the outer core of neutron stars where the main component is neutrons. For example, in the crustal region neutrons are considered to exhibit superfluidity by associating Cooper pairs in the band structure generated by crystalized atomic nuclei. Dynaical structure changes in the crustal region will depend on the EOS of the neutron matter \cite{11}. The EOS also gives the physical properties of nuclei including protons and neutrons. Nuclear matter in which protons and neutrons exist in the same proportion is called symmetric nuclear matter. The dipolar polarizability of symmetric nuclear matter is determined by the symmetric energy, which is the energy difference between neutron matter and symmetric nuclear matter at a given density \cite{12}. For neutron-rich nuclei, the density distributions of protons and neutrons are determined by the balance between two forces. One force is the attractive force due to the symmetric energy working to reduce the density imbalance between protons and neutrons. The other force is a repulsive force to decrease the neutron density to lower the Fermi energy. In this way, the skin depth of pure neutrons on the surface of neutron-rich nuclei is determined by reducing the total energy \cite{13}. 
 
If exact ab initio calculations for an arbitrary many-body quantum system were possible, we could determine the inner structure of neutron stars directly. However, it is impossible in principal to make these calculations using classical computers due to the huge calculation cost \cite{14}. While a method of quantum Monte Carlo calculations is expected to derive values close to exact solutions, it has a negative sign problem for Fermi systems at present, which decreases calculation accuracy significantly \cite{15}. Generally, in a strongly correlated many-body quantum system, where the kinetic energy and the interaction energy are of the same order, theoretical models approximated appropriately are required. However, the validity of the theoretical models must be examined. For hadron matter, the simplest particle system is neutron matter. Therefore, it is important to develop a theoretical model for neutron matter to progress to more complicated hadron systems. If we could experimentally prepare homogeneous neutron matter at various densities and could measure their physical properties without relying on a theoretical model, we could examine the theoretical models using the experimental data. However, it is not strictly necessary to use neutrons to study neutron matter. Using other suitable physical systems which show the same physical quantities as neutron matter, we can simulate neutron matter and study the physics, namely cold atom quantum simulators.

In this review paper, we introduce studies toward the determination of the EOS of neutron matter using a cold atom quantum simulator. The remainder of the paper is organized as follows. In Section \ref{Sec2} we discuss the universal physics of spin-1/2 Fermi particles interacting via s-wave scattering. In Section \ref{Sec3} we explain the principle of quantum simulation of dilute neutron matter using ultracold atoms. In Section \ref{Sec4} we briefly introduce our experimental method. In Section \ref{Sec5} we present the universal EOS of a Fermi system determined by the quantum simulator. Finally, we present the EOS of dilute neutrons obtained from the determined universal EOS.

\section{Universal equation of state for spin-1/2 fermions interacting via s-wave scattering \label{Sec2}}

\begin{table}[pt]
\tbl{Comparison of dilute neutron matter and ultracold $^6$Li atoms.\label{Table}}
{\begin{tabular}{@{}lcc@{}} \toprule
 & Dilute neutron matter & Ultracold $^6$Li atoms \\ \colrule
Mass	 & 940~MeV & 5600~MeV\\
Inter-particle force & nuclear force & electromagnetic force\\
Interaction range ($r_0$) & $\sim$ 2~fm & $\sim$ 2~nm\\
Inter-particle distance ($d=n^{-1/3}$) & $>$ 2.3~fm ($n=n_0/2$) & $\sim$ 250~nm\\
Temperature ($T$)  & 10$^5$--10$^7$~K & $\sim$ 100~nK \\
Thermal length $\left(\lambda_T=\frac{\hbar}{\sqrt{2\pi mk_B T}}\right)$ & 100--900~fm&$\sim$350~nm\\
s-wave scattering length ($a_s$)& $-$18.9~fm & $-\infty$--$+\infty$ \\
Effective range ($r_e$) & 2.75~fm & 4.7~nm \\ \colrule
Potential type & short range&short range\\
Dimensionless temperature ($T/T_F$) & $\sim 0$ & $>$ 0.06 \\
Dimensionless scattering length ($1/k_Fa$) & $-\infty$--$-0.04$ ($0$--$n_0/2$) & $-\infty$--$+\infty$\\
Dimensionless effective range ($k_F r_e$)  & $0$--$3.7$ ($0$--$n_0/2$) & $\sim$ 0.05 \\ \botrule
\end{tabular}}
\begin{tabnote}
nuclear saturation density: $n_0=0.16$~fm$^{-3}$ 
\end{tabnote}
\end{table}

In this section, we provide an overview of the fundamental physics of spin-1/2 fermions interacting via s-wave scattering.
We understand that scattering length and effective range give universal EOS for the fermions, that does not depend on the particle size or mass, the details of the interaction potential, or the absolute value of the energy scale of the particle system.

To clarify the characteristics of neutron matter, the parameters are summarized in Table \ref{Table}. The shown temperature is the typical temperature range of neutron stars. In the dilute density region ($n<n_0/2$) where the average inter-particle distance $d$ is larger than the interaction range $r_0$ of the nuclear force, the neutron matter satisfies the hierarchical structure of the following length scale:
\begin{equation}
r_0 < d\sim k_F, \lambda_T.
\label{eq1}
\end{equation}
Here, $k_F=(3\pi^2n)^{1/3}$ is the Fermi wave number. This inequality corresponds to the condition that the particles undergo low energy scattering with respect to the interaction potential.

Under the condition of dilute and low energy in Eq. (\ref{eq1}), fermions behave as matter waves rather than particles. For two particles with the same mass $m$ interacting with each other by a central potential $U(r)$, where $r$ is the inter-particle distance,
their relative motion can be thought as the motion of one particle with a reduced mass of $ m_r = m / 2 $ following the Schr\"odinger equation:
\begin{equation}
\left[ \frac{1}{2m_r}\left( p_r^2+\frac{{\bf L}^2}{r^2}\right)+U(r) \right]\Psi(r,\theta,\phi)=E\Psi(r,\theta,\phi).
\label{eq2}
\end{equation}
The general solution is given by $\Psi(r,\theta,\phi)=\psi _l(r)Y_{lm}(\theta,\phi)$. Here, $Y_{lm}(\theta,\phi)$ is a spherical harmonic function, $l$ is the bearing quantum number, and $m$ is the magnetic quantum number. By separating Eq. (\ref{eq2}) into radial and angular components, the radial Schr\"odinger equation is given by the following equation:
\begin{equation}
\left[ \frac{\hbar^2}{2m_r}\left( -\frac{d^2}{dr^2}-\frac{2}{r}\frac{d}{dr}+\frac{l(l+1)}{r^2} \right)+U(r) \right]\psi _l(r)=E\psi _l(r).
\label{eq3}
\end{equation}
For a short-range potential, the interaction potential in Eq. (\ref{eq3}) can be neglected for large interaction ranges ($r\gg r_0$):
\begin{equation}
\left[ \frac{\hbar^2}{2m_r}\left( -\frac{d^2}{dr^2}-\frac{2}{r}\frac{d}{dr}+\frac{l(l+1)}{r^2} \right) \right]\psi _l(r)=E\psi _l(r).
\label{eq4}
\end{equation}
The general solution is $\psi _l(k,r)=\frac{c_l}{r}{\rm sin}\left(kr+\eta _l -\frac{1}{2}l\pi \right)$. Here, $c_l$ and $\eta _l$ are the normalization constant and the phase shift of each partial wave. The effect of the short-range interaction appears in the phase shift $\eta _l$ of the scattered wave. In particular, for s-wave scattering of the lowest order ($l=0$), the scattered wave is given by
\begin{equation}
\psi(r,k)\propto \frac{1}{kr}{\rm sin}(kr+\eta_0),
\label{eq5}
\end{equation}
where $\eta_0$ is the phase shift due to s-wave scattering.

The phase shift caused by s-wave scattering can be approximated by the following expansion with the two parameters of scattering length $ a_s $ and effective range $ r_e $:
\begin{equation}
{\rm cot}\eta_0=-\frac{1}{a_sk}+\frac{1}{2}r_ek.
\label{eq6}
\end{equation}
The scattering length gives the phase shift at the long wavelength limit ($k\rightarrow 0$). The effective range gives a correction for finite wave numbers. At the long wavelength limit, the scattering length describes the behavior of the wave function approaching the scattering potential. This definition for the s-wave scattering length is known as the Bethe--Peierls boundary condition \cite{16}. While the scattering length and the effective range have dimensions of length, they do not represent the interaction range. Two particles will only interact when they approach within $r<r_0$.

Next, we consider the grand-canonical many-body Hamiltonian $ \hat{\mathcal{H}}-\mu \hat{\mathcal{N}}$ for spin-1/2 fermions interacting with the s-wave scattering length and the effective range. Let the $n$th eigenstate be $\left. |\Psi_n \right>$, and $K_n=\left< \Psi_n \left| \hat{\mathcal{H}}-\mu \hat{\mathcal{N}} \right| \Psi_n \right>$ be its eigenvalue.
The eigenvalue $K_0$ for the ground state is equal to the thermodynamic potential $\Omega$ at zero temperature ($T = 0$).
At finite temperature ($T>0$), the thermodynamic potential is given by
\begin{equation}
\Omega=-k_B T \ln Z_G,
\label{eq8}
\end{equation}
where the grand partition function is given by
\begin{equation}
Z_G=\sum_n e^{-\frac{K_n}{k_B T}}.
\label{eq7}
\end{equation}

In the case of a system of spin-1/2 fermions interacting via s-wave scattering, three or more body interactions are excluded by Pauli's exclusion rule.
Therefore, the Hamiltonian contains only one-particle kinetic energies and two-particle interaction energies.
While the two-particle interaction potential is necessary, we do not have to prepare the exact interaction potential in the Hamiltonian, because only the phase shift is important for s-wave scattering particles.
An arbitrary interaction potential that can reproduce the scattering length $a_s$ and effective range $r_e$ produces the same physics.
Therefore, we can prepare an artificial interaction potential which reproduces the scattering phase shift by using two parameters of $a_s$ and $r_e$. Such an artificial interaction potential is called a pseudo-potential.
Consequently, the eigenvalues $K_n$ and the grand partition $Z_G$ function depend on the two parameters.
From Eqs. (\ref{eq8}) and (\ref{eq7}), we can see that the thermodynamic potential of the system is given by five thermodynamic variables as
\begin{equation}
\Omega =\Omega(V,T,\mu,a_s^{-1},r_e).
\label{eq9}
\end{equation}
For convenience, we use an inverse scattering length.

When the Hamiltonian has an arbitrary parameter $\lambda$, the following Hermann--Feynman theorem holds:
\begin{equation}
\frac{\partial K_n(\lambda)}{\partial \lambda}=\left< \Psi_n \left| \frac{\partial ( \hat{\mathcal{H}}(\lambda)-\mu \hat{\mathcal{N}} ) }{\partial \lambda} \right| \Psi_n \right>=\left< \Psi_n \left| \frac{\partial  \hat{\mathcal{H}}(\lambda)  }{\partial \lambda} \right| \Psi_n \right>.
\label{eq10}
\end{equation}
Using this theorem and Eq. (\ref{eq8}), the change in the thermodynamic potential with respect to the change in $\lambda$ is given by the following equation:
\begin{eqnarray}
\left( \frac{\partial \Omega}{\partial \lambda}\right)_{V,T,\mu} &=& \frac{\sum_n \frac{\partial K_n(\lambda)}{\partial \lambda}\exp \left[-\frac{K_n}{k_BT}\right]}{\sum_n \exp \left[-\frac{K_n}{k_BT}\right]}\nonumber\\
&=&\frac{\sum_n \left< \Psi_n \left| \frac{\partial  \hat{\mathcal{H}}(\lambda) }{\partial \lambda} \right| \Psi_n \right>\exp \left[-\frac{K_n}{k_BT}\right]}{\sum_n \exp \left[-\frac{K_n}{k_BT}\right]}\nonumber\\
&=& \left< \left< \frac{\partial  \hat{\mathcal{H}}(\lambda)  }{\partial \lambda} \right>\right>.
\label{eq11}
\end{eqnarray}
The outer bracket in the right-hand side of the last line of Eq. (\ref{eq11}) means the thermal mean value by the Boltzmann factor $\exp \left[-\frac{K_n}{k_BT}\right]$ in the many-body system. Now, the Hamiltonian has two parameters, $a^{-1}$ and $r_e$. It is theoretically derived that a change in the thermodynamic potential by changing each parameter follows the following formulae \cite{17}:
\begin{eqnarray}
\left( \frac{\partial \Omega}{\partial a_s^{-1}}\right)_{V,T,\mu,r_e} &=&-\frac{\hbar^2}{4\pi m}C,\label{eq12}\\
\left( \frac{\partial \Omega}{\partial r_e}\right)_{V,T,\mu,a^{-1}} &=&\frac{\hbar^2}{16\pi m}D.\label{eq13}
\end{eqnarray}
Here, $C$ and $D$ are called the contact \cite{18,19,20,21,22} and derivative contact \cite{17}, respectively, and they are extensive physical quantities, whereas the scattering length and effective range are intensive physical quantities. These relations show that the scattering length and effective range are not only two-body scattering parameters but also thermodynamic variables in the many-body system.

Using Eqs. (\ref{eq12}) and (\ref{eq13}), we can add the influence of s-wave scattering to the general total derivative of the thermodynamic potential:
\begin{equation}
d\Omega=-PdV-SdT-Nd\mu-\left( \frac{\hbar^2}{4\pi m}C \right)da_s^{-1}+\left( \frac{\hbar^2}{16\pi m}D \right)dr_e.
\label{eq14}
\end{equation}
Here, $P$, $S$, and $N$ are pressure, entropy, and total number of particles, respectively.
Since $\Omega(\lambda V,T,\mu,a_s^{-1},r_e)=\lambda \Omega(V,T,\mu,a_s^{-1},r_e)$ holds for any volume change ($V\rightarrow \lambda V$),
the following relationship is maintained:
\begin{equation}
\Omega=-PV.
\label{eq15}
\end{equation}
Assuming a homogeneous system, the local thermodynamic quantities are given by $s=S/V$, $n=N/V$, $\mathcal{C}=C/V$, and $\mathcal{D}=D/V$.
Then, the total differential of the pressure is given by Eqs. (\ref{eq14}) and (\ref{eq15}) as follows:
\begin{equation}
dP=sdT+nd\mu+\left( \frac{\hbar^2}{4\pi m}\mathcal{C} \right)da_s^{-1}-\left( \frac{\hbar^2}{16\pi m}\mathcal{D} \right)dr_e.
\label{eq16}
\end{equation}
This total differential equation corresponds to the Gibbs-Duem equation.
By setting the free energy density to $\mathcal{F}=F/V$, it is straightforward to derive the total differential as
\begin{equation}
d\mathcal{F}=-sdT+\mu dn-\left( \frac{\hbar^2}{4\pi m}\mathcal{C} \right)da_s^{-1}+\left( \frac{\hbar^2}{16\pi m}\mathcal{D} \right)dr_e,
\label{eq17}
\end{equation}
by using the standard relation $\Omega=F-\mu N$ and Eq. (\ref{eq15}).
The thermodynamic potential per unit volume can then be described by
\begin{equation}
P=P(T,\mu,a_s^{-1},r_e)
\label{eq18}
\end{equation}
for the grand-canonical ensemble.
The free energy density can be described by
\begin{equation}
\mathcal{F}=\mathcal{F}(T,n,a_s^{-1},r_e)
\label{eq19}
\end{equation}
for the canonical ensemble.

Next, we derive the thermodynamic relation between pressure $P$ and internal energy density $\mathcal{E}$ ($=E/V$) using Eqs. (\ref{eq17}) and (\ref{eq19}).
The dimensions of length and time in the free energy density and the thermodynamic variables are, respectively, $\mathcal{F}$[L$^{-1}$T$^-2$], $T$[L$^{2}$T$^{-2}$], $n$[L$^{-3}$], $a_s^{-1}$[L$^{-1}$], and $r_e$[L].
When the length scale is multiplied by $\lambda$ and the time scale is multiplied by $\lambda^2$, the EOS expressed in Eq. (\ref{eq19}) can satisfy the following scale invariance:
\begin{equation}
\mathcal{F}\left( \frac{T}{\lambda^2},\frac{n}{\lambda^3},\frac{a_s^{-1}}{\lambda},\lambda r_e \right)=\frac{1}{\lambda^5}\mathcal{F}(T,n,a_s^{-1},r_e).
\label{eq20}
\end{equation}
By differentiating both sides with $\lambda$ and using Eq. (\ref{eq17}) with $\lambda=1$, the relation $2Ts-3n\mu+\frac{\hbar^2}{4\pi m a_s}\mathcal{C}+\frac{\hbar^2 r_e}{16\pi m}\mathcal{D}=-5\mathcal{F}$ is obtained.
After a simple derivation with the standard thermodynamic relations $\mathcal{F}=\mathcal{E}-Ts$ and $E=TS-PV+\mu N$, we obtain the simple pressure-energy relation \cite{17,19,21,22}:
\begin{equation}
P=\frac{2}{3}\mathcal{E}+\frac{\hbar^2}{12\pi ma_s}\mathcal{C}+\frac{\hbar^2 r_e}{48\pi m}\mathcal{D}.
\label{eq21}
\end{equation}
This relation holds at all temperatures where s-wave scattering holds. Also, it holds regardless of whether the Fermi system is in the normal state or in the superfluid state.

Below, we consider the dimensionless EOS normalized by a reference length scale and the energy scale of the particle system.
For the grand-canonical ensemble $ P (T, \ mu, a_s ^ {- 1}, r_e) $, we can choose one of the thermodynamic quantities as the reference physical quantity of the particle system.
If we choose temperature $T$ as the reference scale, the energy scale is $k_B T$ and the length scale is $\lambda_T (T) = \sqrt {\frac{2\pi\hbar^2}{mk_BT}}$.
In this case, the EOS can be described in the following dimensionless form:
\begin{equation}
P \frac{\lambda_T^3(T)}{k_BT}=f_P\left( \frac{\mu}{k_BT}, \frac{\lambda_T(T)}{a_s}, \frac{r_e}{\lambda_T(T)} \right).
\label{eq22}
\end{equation}
The left-hand side is dimensionless pressure and the right-hand side $f_P$ is the dimensionless EOS as a function of three dimensionless parameters.
This EOS clearly shows that the dimensionless pressure is given by the universal EOS $f_P$, with no dependence on the details of the particle, the absolute energy scale, and the absolute length scale.
The important parameters for characterizing the many-body system are the three dimensionless parameters given by $\frac{\mu}{k_BT}$, $\frac{\lambda_T}{a_s}$, and $\frac{r_e}{\lambda_T}$.

At zero temperature, we cannot choose $T$ as the reference scale.
In this case, it is natural to choose $\mu$ as the reference energy scale, and choose the inverse of $k_{\mu}(\mu)=\frac{\sqrt{2m\mu}}{\hbar}$ as the reference length scale.
Using these reference scales, we can prepare the dimensionless pressure as $\frac{P}{\mu k_{\mu}^3(\mu)}$.
Since the pressure of non-interacting spin-1/2 fermions at zero temperature is $P_0(\mu)=\frac{2}{15\pi^2}\mu k_{\mu}^3(\mu)$, it is practical to define the dimensionless pressure as
\begin{equation}
\frac{P}{P_0(\mu)}=f_P\left( \frac{1}{k_{\mu}(\mu)a_s}, k_{\mu}(\mu)r_e \right).
\label{eq23}
\end{equation}
This dimensionless EOS $f_P$ corresponds to the ratio of deviation from the EOS of the ideal Fermi gas due to the s-wave interaction.
In this case, the two dimensionless parameters given by $\frac{1}{k_{\mu}a_s}$ and $k_{\mu}r_e$ characterize the many-body system.

When we chose the particle number density $n$ as the reference physical quantities for the canonical ensemble $\mathcal{F}(T,n,a^{-1})$, it is natural to choose the inverse Fermi wave number, $k_{F}(n)=\frac{\sqrt{2m\varepsilon_{\rm F}(n)}}{\hbar}$, instead of the inter-particle distance $d=n^{-1/3}$.
Then, the corresponding reference energy scale is the Fermi energy $\varepsilon_{\rm F}(n)=\frac{\hbar^2}{2m}(3\pi^2n)^{2/3}$.
Using these reference scales, we can prepare the dimensionless free energy density as $\frac{\mathcal{F}}{n \varepsilon_{\rm F}(n)}$.
Since the internal energy density of non-interacting spin-1/2 fermions at zero temperature is $\mathcal{E}_0=\frac{3}{5}n \varepsilon_{\rm F}(n)$, we define the dimensionless free energy density as
\begin{equation}
\frac{\mathcal{F}}{\mathcal{E}_0(n)}=f_\mathcal{F}\left( \frac{T}{T_{\rm F}(n)}, \frac{1}{k_{\rm F}(n) a_s},k_F(n)r_e \right).
\label{eq24}
\end{equation}
In this case, the three dimensionless parameters given by $\frac{T}{T_F}$, $\frac{1}{k_{\rm F} a_s}$, and $k_Fr_e$ characterize the many-body system.

\begin{figure}[th]
\centerline{\includegraphics[width=10cm]{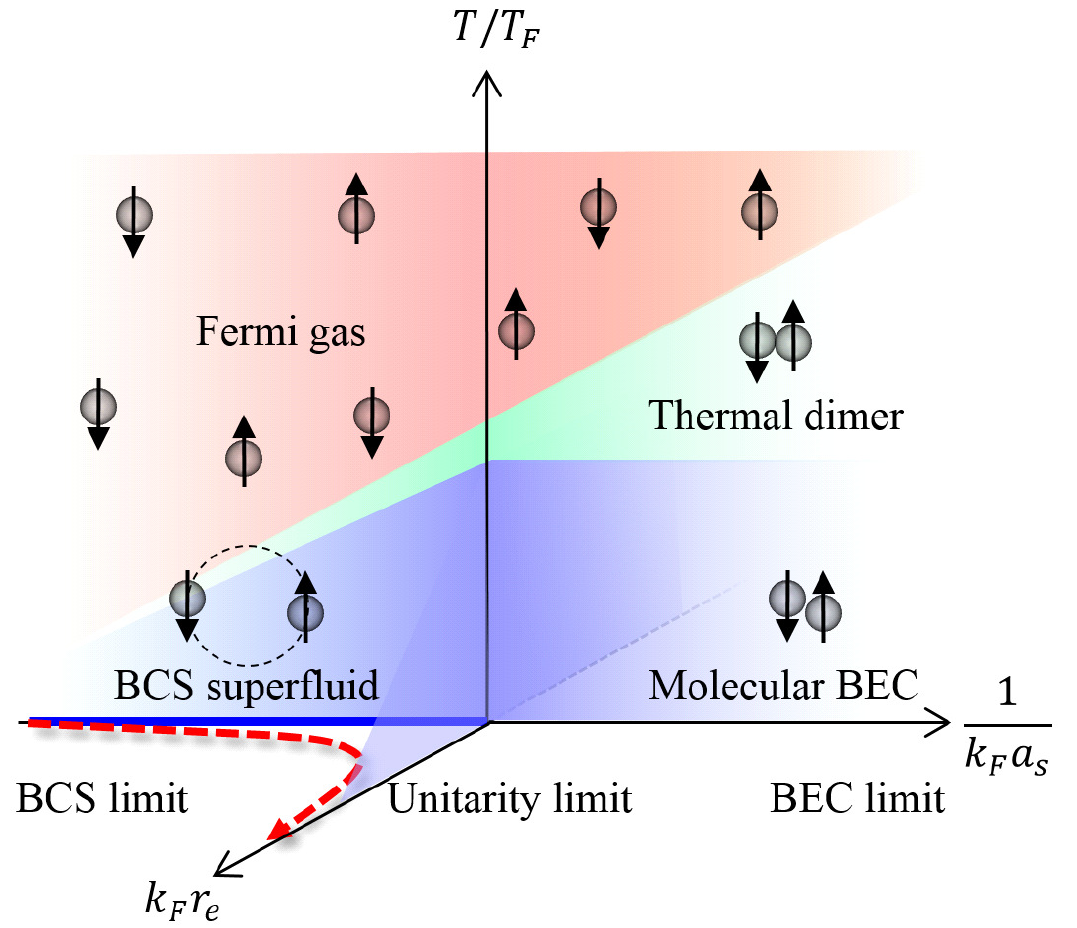}}
\caption{(color online)
Conceptual drawing of the universal phase diagram of the free energy density for a system of spin-1/2 fermions interacting via s-wave scattering.
The red area, green area, and blue area indicate the phases of Fermi gas, thermal dimer, and superfluid of paired fermions, respectively.
The thick blue line in the BCS region at $T/T_F=0$ shows the area where the EOS is determined by the present cold-atom quantum simulator at the zero-range limit ($k_Fr_e=0$).
The red dotted arrow shows how the interaction region changes when the density of neutron matter changes from dilute to dense according to the scattering length and the effective range, as shown in Fig. \ref{Fig2}(b).
}
\label{Fig1}
\end{figure}

Fig. \ref{Fig1} shows a conceptual drawing of the universal phase diagram of the free energy density for a system of spin-1/2 fermions interacting via s-wave scattering.
At the zero-range limit ($k_Fr_e=0$), there is a universal two-body bound state in a vacuum in the region of $a_s>0$.
The wave function and the binding energy are determined by the s-wave scattering length, and they are given by $\psi_b( r = | {\bf r}_\uparrow-{\bf r}_\downarrow |)=\frac{1}{\sqrt{2\pi a_s}} \frac{e^{-r/a_s}}{r}$ and $E_b(a_S)=-\frac{\hbar^2}{ma_s^2}$, respectively.
The interaction region of $1/a_s>0$ at $r_e=0$ is called the strong-coupling region, or the Bose--Einstein condensation (BEC) region, because two fermions with opposite spins can form a bosonic dimer in a vacuum at low temperature, and the bosons can realize a BEC below the critical temperature.
The interaction region of $1/a_s<0$ at $r_e=0$ is called the weak-coupling region, or the BCS region, because the attractive interaction in not sufficient to form dimers, though the fermions can realize Bardeen--Cooper--Schrieffer (BCS) superfluidity below the critical temperature by forming cooper pairs thanks to the many-body effect.
The intermediate interaction point at $1/a_s=0$ between the BCS region and the BEC region is called the unitarity limit.
At the zero temperature limit ($T/T_F=0$), the ground state changes smoothly from BCS superfluidity to the dimer BEC without a phase transition by changing the value of $1/k_Fa_s$.
Such a continuous change of the many-body ground state is called the BCS--BEC crossover \cite{23,24,25,26,27}.

The influence of the effective range is expected to be to strengthen the attractive interaction for $k_Fr_e<0$ and weaken the attractive interaction for $k_Fr_e>0$, because the scattering length and the effective range have opposite signs in the equation giving the phase shift, as shown in Eq. (\ref{eq1}).
For example, we introduce the influence of the effective range in the EOS at the high temperature region.
At high temperatures, the EOS can be expanded with the fugacity $\exp(\mu/k_BT)$ with virial coefficients.
The second order virial coefficient, which is the leading term, reflects the magnitude of the attractive interaction between the two particles.
As shown in Eq. 40 in the previous theoretical work Ref.~\refcite{28}, the effective range works in the opposite direction to the scattering length for interactions.
Therefore, we can expect that the system changes to the strong-coupling region for $k_Fr_e\rightarrow-\infty$ and to the weak-coupling region for $k_Fr_e\rightarrow+\infty$.

The expression for the EOS for the free energy density shown in Eq. (\ref{eq24}) applies even at $T=0$.
Since $\mathcal{E}=\mathcal{F}$ at $T=0$,
the dimensionless internal energy density can be given by the following expression:
\begin{equation}
\frac{\mathcal{E}}{\mathcal{E}_0(n)}=f_\mathcal{E}\left(  \frac{1}{k_{\rm F}(n) a_s},k_F(n)r_e \right).
\label{eq25}
\end{equation}
This EOS can be expanded with $\frac{1}{k_Fa_s}$ and $k_Fr_e$ around the unitarity limit using the relation in Eq. (\ref{eq17}):
\begin{equation}
f_\mathcal{E}^{\rm Unitary}=\xi-\frac{5\pi}{2}\frac{\mathcal{C}}{k_F^4}\left(\frac{1}{k_Fa_s}\right)+\frac{5\pi}{8}\frac{\mathcal{D}}{k_F^6}(k_Fr_e).
\label{eq26}
\end{equation}
Here, $f_\mathcal{E}(0, 0)=\xi$ is the universal value at the unitarity limit,
called the Bertsch parameter \cite{29}.
At the weak-coupling limit ($a_s<0, k_F|a_s|\ll1, k_F|r_e|\ll1$), the EOS obeys the following asymptotic behavior \cite{30,31}:
\begin{equation}
f_\mathcal{E}^{\rm BCS}=1+\frac{10}{9\pi}(k_Fa_s)+\frac{4(11-2\rm{ln}2)}{21\pi ^2}(k_Fa_s)^2+\frac{1}{6\pi}(k_Fr_e)(k_Fa_s)^2+0.032(k_Fa_s)^3
\label{eq27}
\end{equation}
Here, the condensation energy caused by the formation of cooper pairs is omitted by assuming $(\frac{5\Delta^2}{8\varepsilon_{\rm F}^2}\ll1)$ \cite{32}.

We note that we assume an infinite system where the system size $L$ is sufficiently larger than the average inter-particle distance ($d\ll L$), in addition to the dilute and the low energy condition in Eq. (\ref{eq1}).
If this condition is not satisfied, we have to consider finite size effects.

\begin{figure}[th]
\centerline{\includegraphics[width=12.5cm]{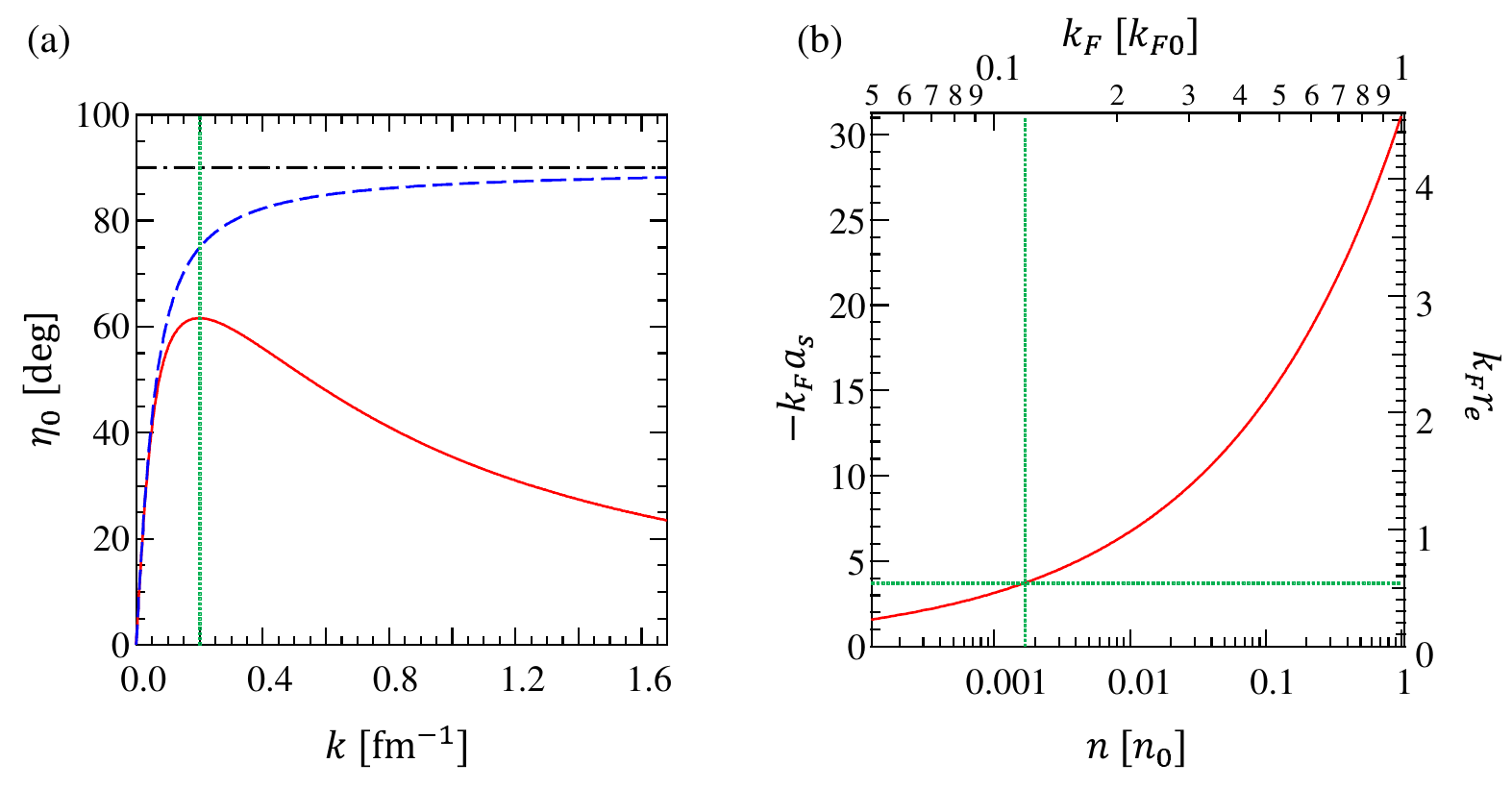}}
\caption{(color online)
s-wave interaction of dilute neutron matter.
(a) s-wave collisional phase shift as a function of wave number.
The red curve shows the phase shift given by Eq. (\ref{eq6}) with neutron parameters of $a_s=-18.5$~fm and $r_e=2.7$~fm.
The blue dashed curve shows the phase shift without the second term of Eq. (\ref{eq6}).
The black dashed indicates the unitary limit ($\eta_0=90$~deg).
The vertical green dotted line indicates the characteristic wave number at $k_c=\sqrt{\frac{2}{|a_s|r_e}}=0.20$ where the phase shift takes the maximum value.
(b) The dimensionless interaction parameters as a function of the neutron density.
The lower horizontal axis is the neutron density in units of the nuclear saturation density of $n_0=0.16$~fm$^{-3}$.
The upper horizontal axis is the Fermi wave number determined by the density in units of the Fermi wave number at the nuclear saturation density as $k_{F0}=(3\pi^2n_0)^{1/3}=1.68$~fm$^{-1}$.
The vertical and the horizontal green dotted lines indicate the density at $k_F(n_c)=k_c$, where $n_c=1.7\times10^{-3}n_0$, $(-k_Fa_s)_c=3.7$ or $(1/k_Fa_s)_c=-0.27$, and $(k_Fr_e)_c=0.54$.
}
\label{Fig2}
\end{figure}

In order to see the dimensionless thermodynamic parameters $1/k_Fa_s$, $k_Fr_e$, and $T/T_F$ as a function of neutron density,
Fig. \ref{Fig2}(a) shows the s-wave collisional phase shift as a function of wave number and Fig. \ref{Fig2}(b) shows $-k_Fa_s$ and $k_Fr_e$ as a function of the neutron density.
The red curve in Fig. \ref{Fig2}(a) shows the phase shift given by Eq. (\ref{eq6}) with the neutron parameters $a_s=-18.5$~fm and $r_e=2.7$~fm.
This curve has a maximum value at the characteristic wave number given by $k_c=\sqrt{\frac{2}{|a_s|r_e}}=0.20$ fm$^{-1}=0.12k_{F0}$, and decreases at wave numbers higher than $k_c$ due to the effect of the finite positive effective range.
Here, $k_{F0}=(3\pi^2n_0)^{1/3}=1.68$~fm$^{-1}$ is the Fermi wave number at the nuclear saturation density.
For comparison, we show the change in the phase shift with zero effective range by the blue dashed curve, and this curve approaches the unitarity limit at $\eta_0=\pi/2$.
Here we define the characteristic density as $k_F(n_c)=k_c$, and it is $n_c=2.7\times 10^{-4}$ fm$^{-3}=1.7\times10^{-3}n_0$ for neutron matter.
Then, Fig. \ref{Fig2}(a) qualitatively shows that the influence of the effective range is small in the dilute density region of $n<n_c$, and the influence of the effective range becomes dominant in the region $n_c<n<n_0/2 $.
Fig. \ref{Fig2}(b) shows the change in $-k_Fa_s$ and $k_Fr_e$ as a function of the neutron density.
The lower horizontal axis is the number density and the upper horizontal axis is the Fermi wave number given by the density.
The left vertical axis is $-k_Fa_s$ for convenience instead of $1/k_Fa_s$, and the right axis is $k_Fr_e$.
At the density $n=n_c$, the dimensionless thermodynamic parameters are $(-k_Fa_s)_c=3.7$ (or $(1/k_Fa_s)_c=-0.27$) and $(k_Fr_e)_c=0.54$.
From this graph, we can see that the many-body state of neutron matter changes from the BCS limit at the low-density limit to the unitarity limit with a large effective range parameter at intermediate density around $n\sim n_0/2$.
The dimensionless temperature can be considered as $T/T_F\sim0$, because the Fermi temperature is several orders of magnitude higher than the absolute temperature.
Therefore, the interaction parameters of neutron matter change according to the red dotted arrow in Fig. \ref{Fig1}.

Since the temperature parameter is negligible for neutron matter, the EOS fundamentally obeys the universal EOS given by Eq. (\ref{eq25}) up to the density $n_0/2$, while the higher partial waves exist \cite{69}.
Therefore, the EOS for the internal energy per particle is given by the following expression as a function of density with the neutron s-wave scattering length and the effective range:
\begin{equation}
\frac{E}{N}=\frac{\mathcal{E}}{n}=\frac{3}{5} \varepsilon_{\rm F}(n)f_\mathcal{E}\left(  \frac{1}{k_{\rm F}(n) a_s},k_F(n)r_e \right).
\label{eq28}
\end{equation}
The pressure is given by Eq. (\ref{eq21}), or the thermodynamic relationship
\begin{equation}
P=n^2\frac{d(E/N)}{dn}.
\label{eq29}
\end{equation}
Consequently, we can determine the fundamental EOS of dilute neutron matter by determining the universal EOS of $f_\mathcal{E}\left(  \frac{1}{k_{\rm F}(n) a_s},k_F(n)r_e \right)$.

\section{Quantum simulator of dilute neutron matter using ultracold atoms \label{Sec3}}

Quantum simulation is a mimic experiment for quantum systems that are difficult to directly examine experimentally and theoretically, by using alternative controllable experimental systems which show the same quantum phenomena under the same Hamiltonian \cite{33,34,35}.
There are various particle systems for quantum simulation, for example, cold atomic systems \cite{36}, Rydberg atomic systems \cite{37}, and cold ionic systems \cite{38}.
We can choose a suitable system for the quantum system of interest.
To simulate dilute neutron matter, the alternative particle system must satisfy the condition given in Eq. (\ref{eq1}), and it is reasonable to control the scattering parameter experimentally.
The particle system which can satisfy this requirement is a cold atomic system.

A cold atomic system is a quantum system of an ultracold neutral atomic gas at a temperature below 1~$\mu$K, realized by using laser cooling and evaporative cooling techniques \cite{54}.
In this work, we use ultracold $^6$Li atoms.
To compare neutron matter with a system of ultracold $^6$Li atoms, we list the typical energy scale and length scale in Table \ref{Table}.
We discuss each item below.

\begin{figure}[th]
\centerline{\includegraphics[width=8.5cm]{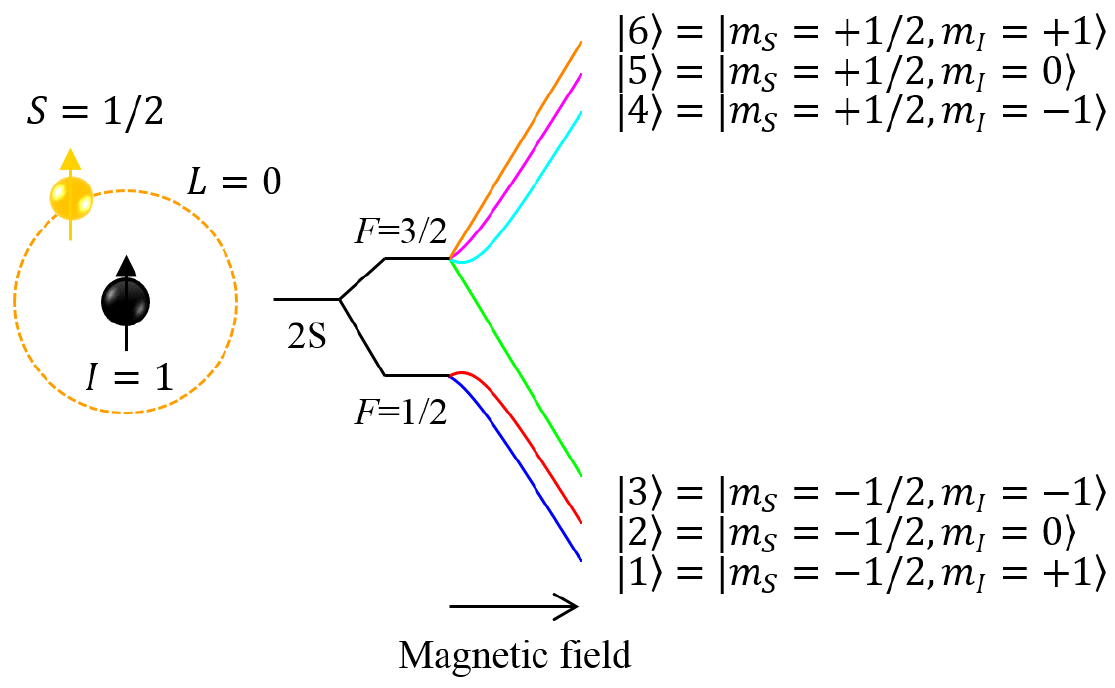}}
\caption{(color online)
Energy diagram of $^6$Li atoms in the electrically ground state in a magnetic field.
The black sphere is the Li nucleus and the yellow sphere is the peripheral 2S electron.
$m_s$ and $m_I$ are the magnetic components of the electric spin and the nuclear spin projected onto the quantum axis.
}
\label{Fig3}
\end{figure}

\paragraph{A spin-1/2 Fermi particle system:}
Figure \ref{Fig3} shows the energy diagram of a $^6$Li atom in the electrical ground state in a magnetic field.
A $^6$Li atom is a Fermi particle since it possesses an odd total number of fermions: three electrons, three protons, and three neutrons.
The mass of a $^6$Li atom is about 6 times that of a neutron.
In the electrical ground state, a $^6$Li atom has a peripheral electron in the 2S orbit, the total electric spin is $S=1/2$, and the orbital angular moment is $L=0$.
Since the nuclear spin is $I=1$, the electrical ground state has two hyperfine levels which have the total angular momenta of $F=1/2$ and $F=3/2$.
Each level splits in a magnetic field due to the Zeeman effect as shown in Fig. \ref{Fig3}.
In a stronger magnetic field than that of the Paschen--Back region, the internal states are divided into the lower branch with $m_s=-1/2$ and the upper branch with $m_s=-1/2$, and each branch has three levels of $m_I= 0,$ $\pm 1$, where $m_s$ and $m_I$ are the magnetic components of the electric spin and the nuclear spin projected onto the quantum axis.
We denote the states in ascending energy as $\ket{1}$, $\ket{2}$, $\dots$, for convenience.
The states $\ket{1}$ and $\ket{2}$ are considered in this work.
Since both these states have longer lifetimes than the experimental time, the particle density and the chemical potential can be held to $n_1=n_2=n/2$ and $\mu_1=\mu_2=\mu$ during experiments when we prepare a balanced mixed state of $\ket{1}$ and $\ket{2}$ as an initial condition.
The energy difference between the two states in the magnetic field is not a factor in the EOS.
In this way, we can realize a stable two-component Fermi system
that is equivalent to a spin-1/2 Fermi system.
Taking the internal state as the spin, the internal states are called pseudo-spins.

\paragraph{Short-range interaction potential:}
Neutral atoms interact with each other by the electromagnetic force through the van der Waals potential.
In the case of neutral atoms in the electric S-orbit, the interaction potential in Eq. (\ref{eq2}) has the asymptotic form $U(r)=-\frac{C_6}{r^6}$ at the large inter-particle distance.
The effective asymptotic potential including the centrifugal potential is given by $U_{\rm eff}(r)=-\frac{C_6}{r^6}+\frac{\hbar^2}{2m_r}\frac{l(l+1)}{r^2}$.
For $^6$Li atoms in the electrical ground state, the coefficient is $C_6=1393.39$~$E_h a_0^6$, where $E_h$ is the Hartree energy \cite{39}.
In the case of p-wave scattering with $l=1$, the interaction potential has a centrifugal potential barrier with a height of about 8~mK.
When the temperature of the $^6$Li atoms is low and the density is dilute enough to realize the condition where the thermal energy and the Fermi energy are much smaller then the potential barrier, the higher partial wave larger than $l=1$ can be suppressed.
Since both the temperature and the Fermi energy in typical cold atom experiments are three orders lower than the p-wave centrifugal potential barrier, we can consider only the s-wave interaction potential of $U(r)=-\frac{C_6}{r^6}$.
Here we define the binding energy of the loosest bound state in the asymptotic s-wave interaction potential as $E_b=-\frac{\hbar^2}{2m_r r_0^2}$.
The length scale of the bound state is $r_0=\left( \frac{2 m_r C_6}{\hbar^2} \right)^{1/4}$ by setting $E_b=U(r_0)$.
This length scale is generally defined as the van der Waals length, and it is $r_0=1.7$~nm for $^6$Li atoms.
As shown in Ref.~\refcite{39}, the wave function approaching the van der Waal potential is modulated in the region $r<r_0$ and its modulation decays exponentially in the region $r>r_0$.
Therefore, we can consider the van der Waals length as the interaction range for the potential.
The mean inter-particle distance is about two orders larger than the interaction range in our experimental conditions.
While it is dilute as a particle, the thermal de Broglie length of the atom reaches the mean distance of the particle.
Therefore, the cold atomic system can be regarded as a quantum system with a remarkable quantum effect.
From the above discussion, we can see that the cold atomic system can satisfy the dilute and low-energy condition given in Eq. (\ref{eq1}).

\begin{figure}[th]
\centerline{\includegraphics[width=12.5cm]{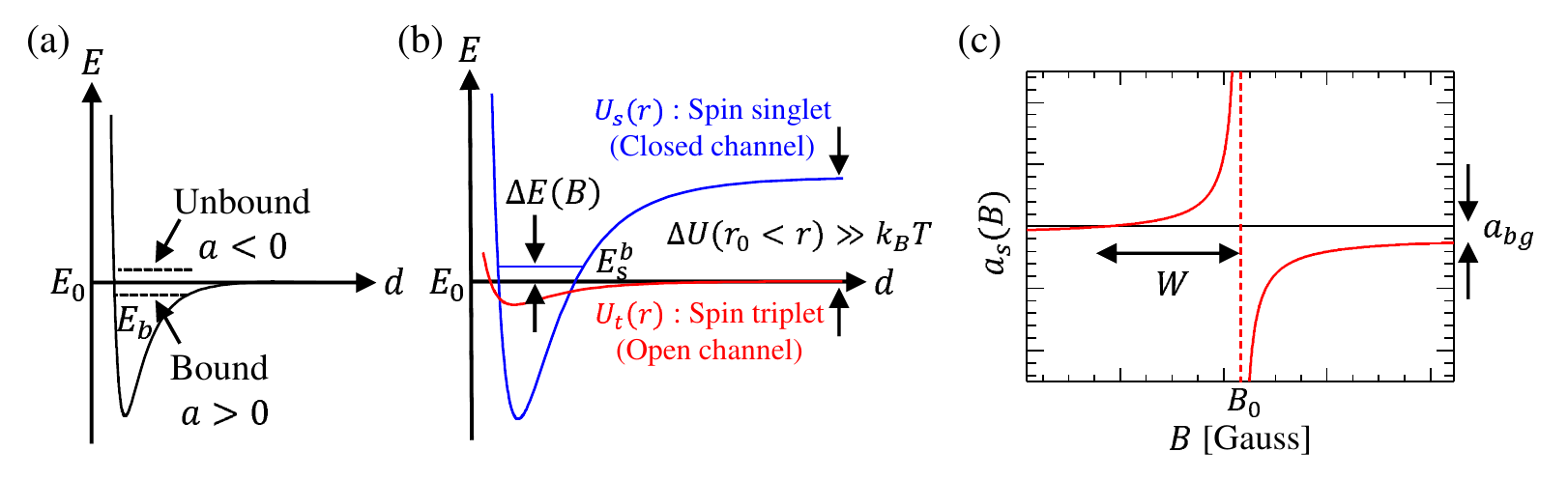}}
\caption{(color online)
Change of the scattering length caused by shape resonance and Feshbach resonance.
(a) Shape resonance.
(b) Feshbach resonance.
(c) An example of a change of the scattering length in the Feshbach resonance.
}
\label{Fig4}
\end{figure}

\paragraph{Control of the s-wave scattering length by using the Feshbach resonance:}
As discussed for Eq. (\ref{eq6}) in Sec. \ref{Sec2}, the s-wave scattering length is determined by the wave function approaching the short-range potential.
The wave function strongly depends on the level of the bound state in the potential because of the boundary conditions, and hence, the scattering length also depends on the level of the bound state.
Fig. \ref{Fig4}(a) shows an arbitrary potential with a bound state located at a level of $E_b$ with respect to the zero energy ($E_0$).
In the case that a bound state exits just below the zero energy ($E_b \lesssim E_0$), the scattering length has a large positive value.
This situation is equivalent to the neutron-proton triplet nuclear force.
It has a large positive scattering length of $a_t \simeq$5.4~fm with respect to the interaction range of $r_0 \simeq$2~fm, and they have a loose bound deuteron state \cite{40}.
On the other hand, when a virtual bound state exits just above the zero energy ($E_b \gtrsim E_0$), the scattering length has a large negative value.
This situation is equivalent to the neutron--proton singlet nuclear force and the neutron--neutron singlet nuclear force.
When the level of the bound state matches the zero energy ($E_b=E_0$), the scattering length diverges.
This is the condition of the unitarity limit.
The manner in which the scattering length varies according to the level of the bound state is called shape resonance \cite{41}.

It is difficult to change the shape of the nuclear force or the van der Waals potential experimentally.
However, if the level of the bound state could be set freely within the scattering potential, we could change the value of the scattering length.
This is the basic idea of Feshbach resonance, proposed by H. Feshbach in 1958 \cite{42} and U. Fano in 1961 \cite{43}.
Here, we briefly explain the principle of the magnetically tunable Feshbach resonance between $^6$Li atoms in $\ket{1}$ and $\ket{2}$, which are defined in Fig. \ref{Fig3}, by using Fig. \ref{Fig4}(b).
In a strong magnetic field, the direction of electric spin of two internal state are the same.
Then, the interaction potential $U_t(r)$ is one of the electric triplet potentials ($S=1, m_S=-1$), where $S$ means the total electric spin.
This potential is called the open channel, where the two scattering particles can enter and exit from the channel.
Next, we consider an electric singlet potential ($S=0, m_S=0$), $U_s(r)$.
The potential depth of $U_s(r)$ is deeper than that of $U_t(r)$, and the energy of $U_s(r)$ is larger than that of $U_t(r)$ at large inter-particle distances in a magnetic field because the internal energy with $m_s=+1/2$ increases in a magnetic field, as shown in Fig. \ref{Fig3}.
The energy difference in a magnetic field is about $\Delta U=\mu_B  g_s B$, where $\mu_B$ and $g_s$ are the Bohr magneton and the electron spin g-factor.
At a $^6$Li Feshbach resonance of around 834~Gauss, the energy difference is about $\Delta U/k_B \sim 100$~mK, which is much higher than the temperature of the gas.
Since it is impossible for particles to enter and exit from the singlet channel due to energy conservation, this singlet channel is called the closed channel.
The level of $U_t(r)$ is sensitive to the magnetic field because it has a magnetic moment. On the other hand, the level of $U_s(r)$ is insensitive to the magnetic field because magnetic moment is tiny.
Therefore, we can control the relative level between the open channel and the closed channel by applying a bias magnetic field to the atoms.

Next, we consider the case where the open channel does not have a bound state, but the closed channel has a bound state $E_s^b$, as shown in Fig. \ref{Fig4}(b).
When the level of $E_s^b$ is tuned to the zero energy $E_0$ of the open channel under the condition where the total angular momentum is conserved between $U_t(r)$ and $U_s(r)$, the wave function entering the open channel is coupled to the bound state in the closed channel.
The coupling strength depends on the detuning given by $\Delta E(B)=E_s^b-E_0(B)$, and the detuning can be tuned by applying a magnetic field.
In this way the s-wave scattering length can be controlled as a function of the magnetic field.
This resonance scattering caused by the other scattering channel is called Feshbach resonance.
In the case of the magnetically tunable Feshbach resonance, the s-wave scattering length obeys the following equation around the resonance magnetic field \cite{39}:
\begin{equation}
a(B)=a_{\rm bg}\left( 1+\frac{W}{B-B_0} \right).
\label{eq31}
\end{equation}
$a_{\rm bg}$, $B_0$, and $W$ are parameters giving the background scattering length, the resonant magnetic field, and the width of the resonance.
They are determined by the details of $U_t(r)$ and $U_s(r)$.
Feshbach resonance was first confirmed using a sodium BEC by an MIT group in 1998 \cite{44}.
Since then, Feshbach resonance has been confirmed for various atomic species and heteronuclear atomic combinations \cite{39}.
Now, a cold atomic system is known to be a unique quantum system with a tunable interaction.
There are two types of Feshbach resonance of $^6$Li atoms \cite{45}.
One is broad Feshbach resonance, where the effective range is almost constant under the Feshbach resonance.
The other is narrow Feshbach resonance, where the effective range also changes as a function of the magnetic field.
In this work, we use a broad Feshbach resonance of $^6$Li between the internal states of $\ket{1}$ and $\ket{2}$.

Our experimental conditions are summarized in Table \ref{Table}.
$^6$Li atoms are cooled down to $T=0.06T_F$ in the superfluid state around the unitarity limit.
The dimensionless parameter for the scattering length ($1/k_Fa_s$) can be tuned to an arbitrary value using Feshbach resonance.
The dimensionless parameter for the effective range ($k_Fr_e$) has a small constant value $ k_Fr_e ~ 0.05 $ over the BCS--BEC crossover, which means the system contributes to the zero range limit.
Therefore, our cold atom quantum simulator can be used to investigate the region indicated by the solid blue line in Fig. \ref{Fig1} for $a_s<0$.

\begin{figure}[th]
\centerline{\includegraphics[width=12.5cm]{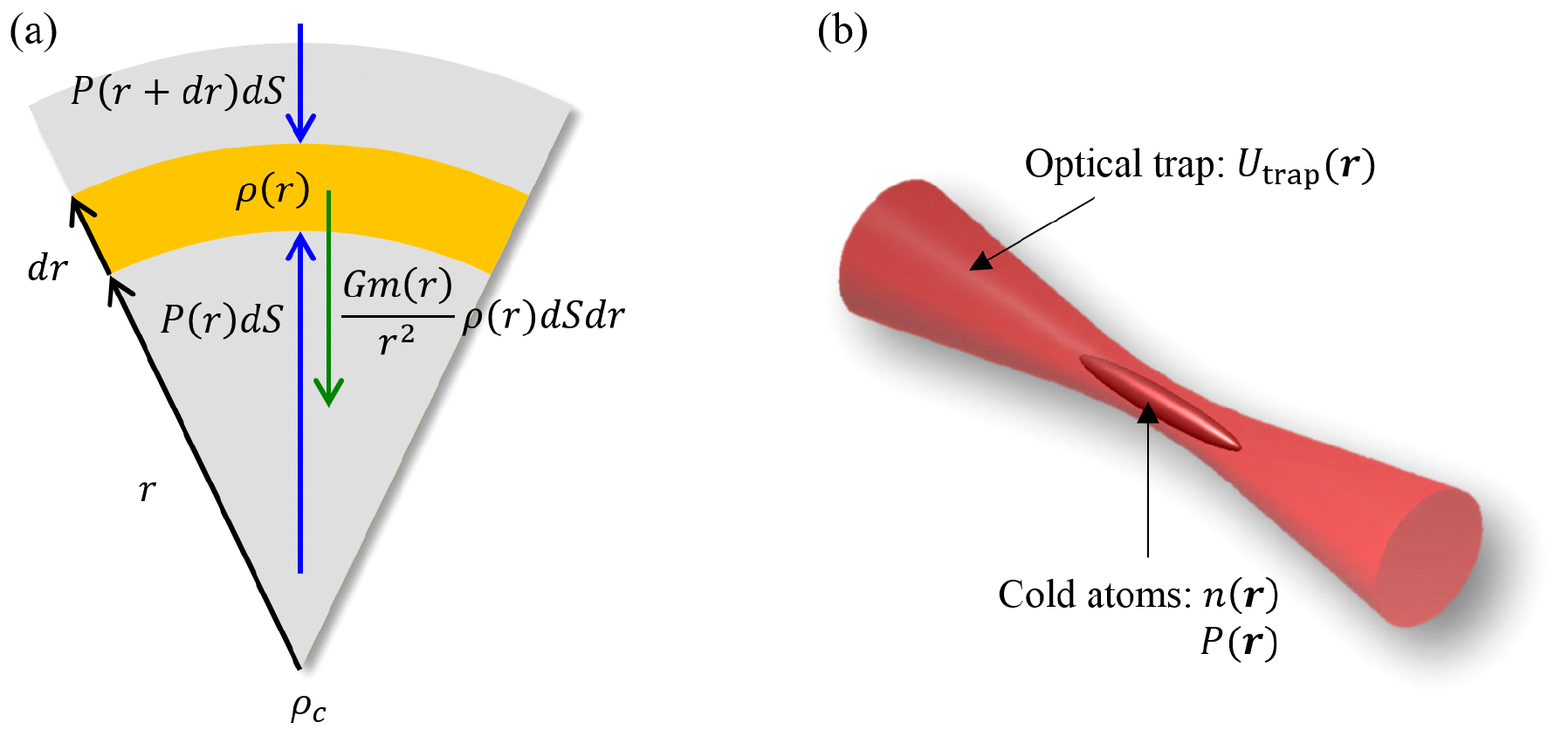}}
\caption{(color online)
Density and pressure distribution determined by the EOS.
(a) Inside a neutron star.
(b) Ultracold atoms trapped in an optical dipole trap.
}
\label{Fig5}
\end{figure}

At the end of this Section, we discuss the similarity of the density and pressure distributions determined by the EOS between the inside of neutron stars and for ultracold atoms in a trap.
In Fig. \ref{Fig5}(a), we show a schematic drawing of a cross section of a neutron star.
Let $M(r)=\int_0^r 4\pi r^2 \rho(r)dr$ be the total mass up to the radius $r$ and let $\rho(r)dSdr$ be the mass in the minute region $dr$.
A stable star without collapse must exhibit a force balance between the pressure of the particles and the gravitational force $F=-\frac{GM(r)}{r^2}\rho(r)dSdr$ at all positions \cite{46}.
This condition is given by
\begin{equation}
\frac{dP(r)}{dr}=-\frac{GM(r)}{r^2}\rho(r).
\label{eq32}
\end{equation}
Here we consider the EOS as $P=P(\rho)$, and we transform the force balance to
\begin{equation}
\frac{d\rho(r)}{dr}=-\frac{GM(r)\rho(r)}{r^2}\left( \frac{dP(\rho)}{d\rho} \right)^{-1}.
\label{eq33}
\end{equation}
This differential equation means that the density distribution inside the star can be uniquely determined by the EOS with the central density $\rho(0)$.
Strictly, the Tolman--Oppenheimer--Volkoff equation including a correction for general relativity to Eq. (\ref{eq33}) is the suitable equation for neutron stars \cite{47,48}.

Next, we consider the case of cold atoms in an optical dipole trap (ODT).
The ODT is a trapping potential produced by the interaction between the electric field of the laser light and the atomic dipole moment \cite{49}.
When the wavelength of the ODT laser is longer than that of the optical transition, the potential depth is proportional to the intensity of the laser light, that is $U_{\rm trap}({\bf r})\propto -I({\bf r})$.
Then we can produce a three-dimensional trapping potential by a focused laser light, as shown in \ref{Fig5}(b).
Under a thermal equilibrium condition in a homogeneous magnetic field for the Feshbach resonance,
the temperature $T$, the scattering length $a_s$, and the effective range $r_e$ have constant values over the trapped system.
Hence, the Gibbs--Duem equation given in Eq. (\ref{eq16}) is simplified to
\begin{equation}
dP({\bf r})=n({\bf r})d\mu.
\label{eq34}
\end{equation}
Here, we introduce a local density approximation (LDA), where we assume that the local thermodynamic quantities are determined by the local density and the influence of the density gradient is negligible, by setting the local chemical potential to
\begin{equation}
\mu({\bf r})=\mu_0-U({\bf r}),
\label{eq35}
\end{equation}
where $\mu_0$ is the chemical potential at the bottom of the trap ($\mu_0=\mu(0)$).
Eq. (\ref{eq35}) gives $d\mu=-\nabla U({\bf r})d{\bf r}$, and then Eq. (\ref{eq34}) is transformed to \cite{50}
\begin{equation}
\nabla P({\bf r})=-n({\bf r})\nabla U({\bf r}).
\label{eq36}
\end{equation}
This equation shows the force balance between the pressure and the central force.
This means that the force balance is equivalent to thermodynamic stability under the LDA.

While it is impossible for a human to fall into a neutron star and see the inside, it is possible to measure the local pressure and local density of cold atoms at each position in a trap potential.
Therefore, we can understand that a quantum simulator using ultracold atoms is suitable for studying dilute neutron matter in terms of a tunable quantum system and observation.

\section{Experimental method \label{Sec4}}

\begin{figure}[th]
\centerline{\includegraphics[width=12.5cm]{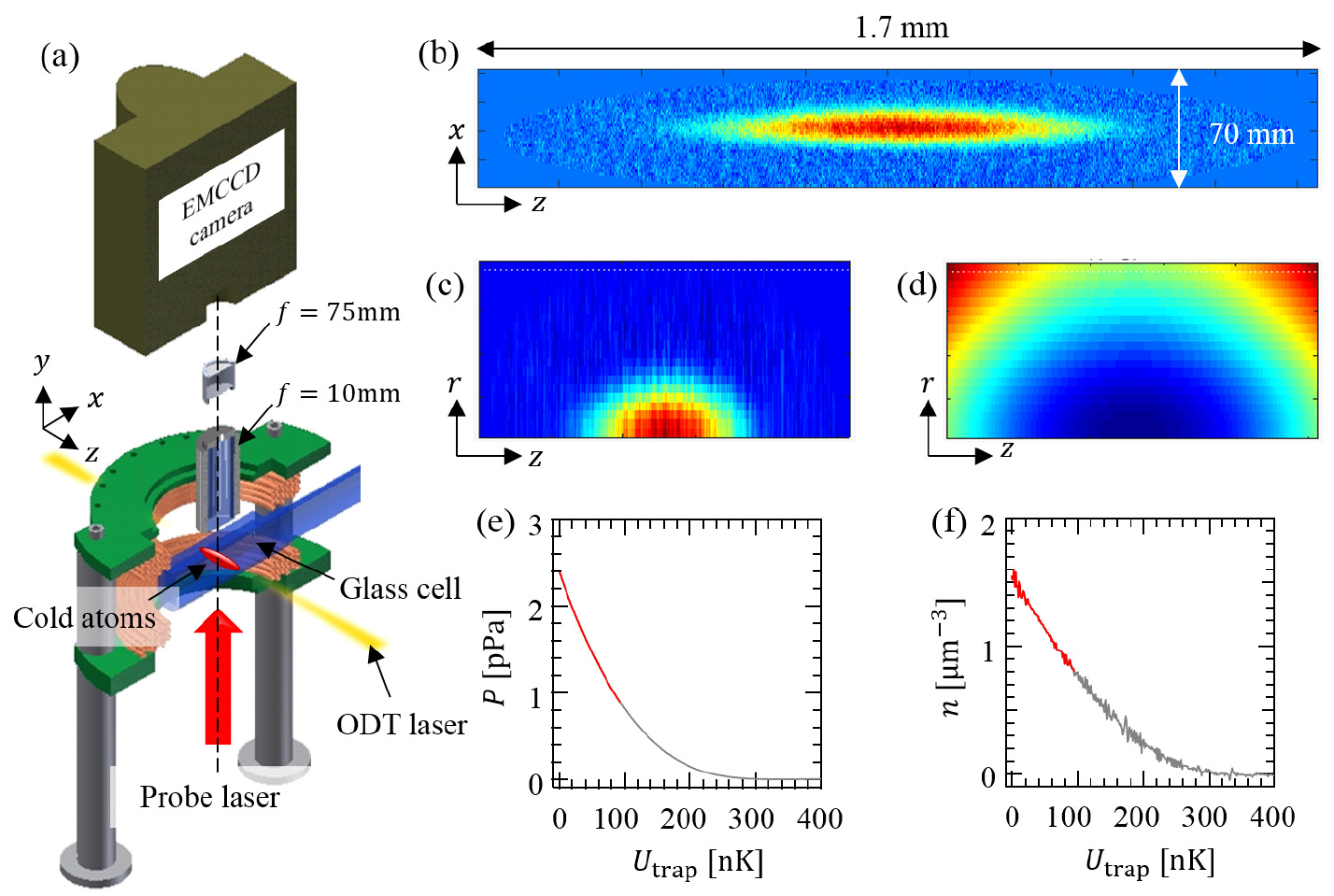}}
\caption{(color online)
Experimental setup and typical experimental data.
(a) Cross section of the experimental apparatus.
The $z$ axis is defined as the direction of the ODT laser light.
(b) An {\it in situ} absorption image $\bar{n}(x,z)$ taken by the CCD camera.
(c) An image of the local pressure $P(r,z)$.
(d) An image of the trapping potential $U_{\rm trap}(r,z)$.
(e) The local pressure as a function of the trapping potential $P(U_{\rm trap})$.
(f) The local number density as a function of the trapping potential $n(U_{\rm trap})$.
The red curves in (e) and (f) show the data region used in the data analysis, where the number density is larger than half of the peak density.
The laser light for laser cooling is not displayed.
}
\label{Fig6}
\end{figure}

In this Section, we overview our experimental method to determine the EOS in the region indicated by the solid blue line in Fig. \ref{Fig1}.
Previous publications can provide more details on the experimental apparatus \cite{51}, observation method \cite{52}, and data analysis \cite{53}.

A cross section of a part of our experimental apparatus is shown in Fig. \ref{Fig6}(a).
Cold atoms are collected in the center of the glass cell using a laser cooling technique \cite{54}.
The outer size of the cell is 30~mm $\times$ 30~mm $\times$ 100~mm, and the glass thickness is 3.5~mm.
In the glass cell, the vacuum is maintained at a level of $10^{-9}$~Pa to suppress heating and loss of trapped atoms by collisions with residual gas.
A pair of coils installed outside the cell produces a quadrupole magnetic field for laser cooling and a bias magnetic field for the Feshbach resonance by controlling the direction of their current.
A microscope objective lens and an eyepiece lens are installed above the cell for imaging the atoms.
This imaging system has a high optical resolution of 2~$\mu$m,
sufficient to observe the density distribution in the trap.
A laser light with a wavelength of 1070~nm is applied to the atoms along the $z$ direction and is focused on the laser cooled atoms with a beam waist of 40~$\mu$m.
The initial laser power of the ODT is 100~W to produce a deep trapping potential of 3~mK to capture the cooled atoms with a temperature of $T\sim 1$mK.
Then, the trapped atoms are further cooled in an evaporative cooling stage by lowering the ODT laser power down to 50~mW.
The final temperature of the $^6$Li atoms reaches 100~nK.
The lifetime of the trapped atoms is about 90~s,
which is long enough to realize the thermal equilibrium condition because the thermalization time of $\sim$100~ms is much shorter than the time constant of the loss process.

The broad Feshbach resonance of $^6$Li atoms between the internal states $\ket{1}$ and $\ket{2}$ has the parameters $a_{\rm bg}=-1582$~$a_0$, $B_0=832.18$~Gauss, and $W=262.3$~Gauss \cite{55}, where $a_0$ is the Bohr radius.
Since $a_{\rm bg}<0$ and $W>0$, the s-wave scattering length changes as shown in Fig. \ref{Fig4}(c).
The Fermi system is in the BEC region for $B<B_0$ and in the BCS region for $B>B_0$.
In this experiment, we cool the trapped fermions by evaporative cooling in the BEC side at 777~Gauss, and produce an almost pure dimer BEC.
Since the dimer BEC accumulates at the bottom of the trap, we can remove the thermal component more efficiently in the BEC region than in the BCS region, where the thermal component and the superfluid component have almost the same density distribution due to Fermi statistics.
After producing the dimer BEC, we sweep the Feshbach magnetic field adiabatically to $B\geq B_0$ to change the Fermi system to the BCS regime.
According to the above procedure, we prepare spin-1/2 fermions at the zero temperature limit with $a_s \leq 0$ in the range $832.18 \leq B <1050$~Gauss.

The density distribution in the trap is observed {\it in situ} by using an absorption imaging technique.
Now, we consider the situation where a resonant probe laser pulse with an incident intensity of $I_{\rm in}(x,z)$ is applied to the atoms along the $y$ axis, as shown in Fig. \ref{Fig6}(a), and the output laser has intensity $I_{\rm out}(x,z)$.
We define the transmittance as $T_{\rm abs}(x,z)=I_{\rm out}(x,z)/I_{\rm in}(x,z)$ and the column density as ${\bar n}(x,z)=\int_{-\infty}^{+\infty}n(x,y,z)dy$.
The  Beer--Lambert  law gives the relation between $T_{\rm abs}(x,z)$ and ${\bar n}(x,z)$ as
\begin{equation}
{\bar n}(x,z)=\frac{1}{\sigma_{\rm abs}}\left[ -{\rm log}(T_{\rm abs}(x,z))+s(x,z)\cdot (1-T_{\rm abs}(x,z)) \right],
\label{eq37}
\end{equation}
where $\sigma_{\rm abs}$ is the absorption cross section, $s(x,z)=I_{\rm in}(x,z)/I_{\rm sat}$ is the saturation parameter, and $I_{\rm sat}$ is the saturation intensity of the optical transition used for imaging.
In the experiment, we measure the intensity distribution of the probe laser pulse using a CCD camera with atoms as $I_{\rm out}(x,z)$ and without atoms as $I_{\rm in}(x,z)$, and use them to calculate the column density by Eq. (\ref{eq37}).
Typical experimental data of the column density is shown in Fig. \ref{Fig6}(b).

It is possible to reconstruct the original three dimensional density distribution $n(r,z)$ from the column density ${\bar n}(x,z)$ integrated along the $y$ axis by an inverse Abel transformation:
\begin{equation}
n(r,z)=-\frac{1}{\pi}\int_r^\infty \frac{1}{\sqrt{x'^2-r^2}}\frac{\partial \bar{n}(x',z)}{\partial x'}dx',
\label{eq38}
\end{equation}
with $r=\sqrt{x^2+y^2}$.
The local pressure can be calculated using Eq. (\ref{eq36}) with the local density distribution $n(r,z)$ and the trapping potential $U_{\rm trap}(r,z)$ as
\begin{equation}
P(r,z)=\int_r^\infty n(r',z)\frac{\partial U_{\rm trap}(r',z)}{\partial r'}dr'.
\label{eq39}
\end{equation}
Even if we integrate along the $z$ direction in this equation, we obtain the same local pressure.

It is also possible to calculate the local pressure directly from the column density.
When Eq. (\ref{eq38}) is substituted for Eq. (\ref{eq39}) and simplified, the following equation is obtained:
\begin{equation}
P(\rho,z)=\frac{1}{\pi}\int_{\rho}^{\infty}dx~{\bar n}(x,z)\left[ \frac{\frac{\partial U_{\rm trap}}{\partial \rho}(x,z)}{(x^2-\rho^2)^{1/2}} + \int_{\rho}^x d\rho '~\frac{\rho '\frac{\partial U_{\rm trap}}{\partial \rho}(x,z)-x\frac{\partial U_{\rm trap}}{\partial \rho}(\rho ',z)}{(x^2-\rho '^2)^{3/2}} \right].
\label{eq40}
\end{equation}
Using this equation, we can calculate the local pressure by a simple integration.
The local density is given by the thermodynamic relation in Eq. (\ref{eq34}) as
\begin{equation}
n(U_{\rm trap})=\frac{dP}{d\mu}=-\frac{dP(U_{\rm trap})}{dU_{\rm trap}}.
\label{eq41}
\end{equation}

Neither of these methods depend on the temperature, the s-wave scattering length, or the effective range.
Thus we can obtain the local density and the pressure accurately from the observed column density and the well known trapping potential.
In this work, we used the latter method.
Fig. \ref{Fig6}(c) shows an image of the local pressure $P(r,z)$ calculated from the column density shown in Fig. \ref{Fig6}(b) by Eq. (\ref{eq40}).
Fig. \ref{Fig6}(d) is an image of the trapping potential $U_{\rm trap}(r,z)$ for the column density.
Fig. \ref{Fig6}(e) shows the local pressure as a function of the trapping potential obtained by relating $P(r,z)$ and $U_{\rm trap}(r,z)$ at each position.
Fig. \ref{Fig6}(e) shows the local density as a function of the trapping potential obtained by Eq. (\ref{eq41}).
We obtained many data sets of pressure and density of atoms with various scattering lengths in this data analysis.

At the unitarity limit, we can determine the EOS from the density and the pressure measured at the unitarity limit at the zero-temperature limit.
The internal energy density is directly given by the energy--pressure relation in Eq. (\ref{eq21}) as
\begin{equation}
\mathcal{E}(n)=\frac{3}{2}P(n).
\label{eq42}
\end{equation}
Since the internal energy density of the non-interacting spin-1/2 fermions, $\mathcal{E}_0(n)$, is given by the density, we can determine the Bertsch parameter by $\xi=\mathcal{E}(n)/\mathcal{E}_0(n)$.
In this work, we determined the Bertsch parameter to be $\xi=0.375(10)$.

Outside of the unitarity limit, it is impossible to use Eq. (\ref{eq21}) to determine the EOS because the contact density $\mathcal{C}$ in the right-hand side has not been determined.
Then, we need to construct the EOS in other ways from the BCS region to the unitarity limit.
For this purpose, we choose $a_s$ as the reference length scale for $P=P(\mu,a_s^{-1})$, while we choose $\mu$ as the reference energy scale in Eq. (\ref{eq23}).
In this case, the reference energy scale becomes $\varepsilon_{a_s}=\frac{\hbar^2}{2ma_s^2}$.
Using these length and energy scales, we prepare the dimensionless chemical potential defined as
\begin{equation}
\frac{\mu}{\varepsilon_{a_s}(a_s)}=\mathcal{G}(\chi).
\label{eq43}
\end{equation}
Here, $\mathcal{G}$ is the universal EOS giving the dimensionless chemical potential as a function of the dimensionless variable $\chi \equiv P\frac{a_s^3}{\varepsilon_{a_s}(a_s)}$.
Under the LDA, $\mathcal{G}$ is given by
\begin{equation}
\frac{\mu_0-U_{trap}}{\varepsilon_{a_s}(a_s)}=\mathcal{G}\left( \chi(U_{trap}) \right),
\label{eq44}
\end{equation}
where $\chi(U_{trap}) \equiv P(U_{\rm trap})\frac{a_s^3}{\varepsilon_{a_s}(a_s)}$.
The dimensionless variable in the right hand side can be determined only from experimental data, and the offset of the left hand side is given by the value of $\mu_0$.
Therefore, we can easily overlap all the experimental data acquired for various scattering length models independently, as shown Fig. 4 in Ref.~\refcite{53}.
This method is inspired by a method used in a previous experiment to determine the EOS of a unitary Fermi gas at finite temperature \cite{58}.
The connected curve has an arbitrary offset of $\Delta \mathcal{G}$ from the true EOS as $\mathcal{G}(\chi)=\mathcal{G}_{\rm true}(\chi)+\Delta \mathcal{G}$.
The possible values of $\Delta \mathcal{G}$ can be constrained by the thermodynamic relation and the EOS at the BCS limit given in Eq. (\ref{eq27}).
From the constructed EOS of $\mathcal{G}(\chi)$, we can convert the EOS to $\frac{P}{P_0}=f_P\left( \frac{1}{k_\mu a_s} \right)$ by the following equations:
\begin{equation}
\begin{cases}
\; \frac{1}{k_\mu a_s}=-\mathcal{G}(\chi)^{-1/2}, \\
\; \frac{P}{P_0}=-\frac{15\pi^2}{2}\chi \mathcal{G}(\chi)^{-5/2}.
\end{cases}
\label{eq45}
\end{equation}
Also, the EOS for the internal energy density $f_\mathcal{E}\left( \frac{1}{k_F a_s} \right)$ can be derived from $f_P\left( \frac{1}{k_\mu a_s} \right)$ by the thermodynamic relations \cite{53}.

\section{Experimental results and the EOS for dilute neutron matter \label{Sec5}}

\begin{figure}[th]
\centerline{\includegraphics[width=12.5cm]{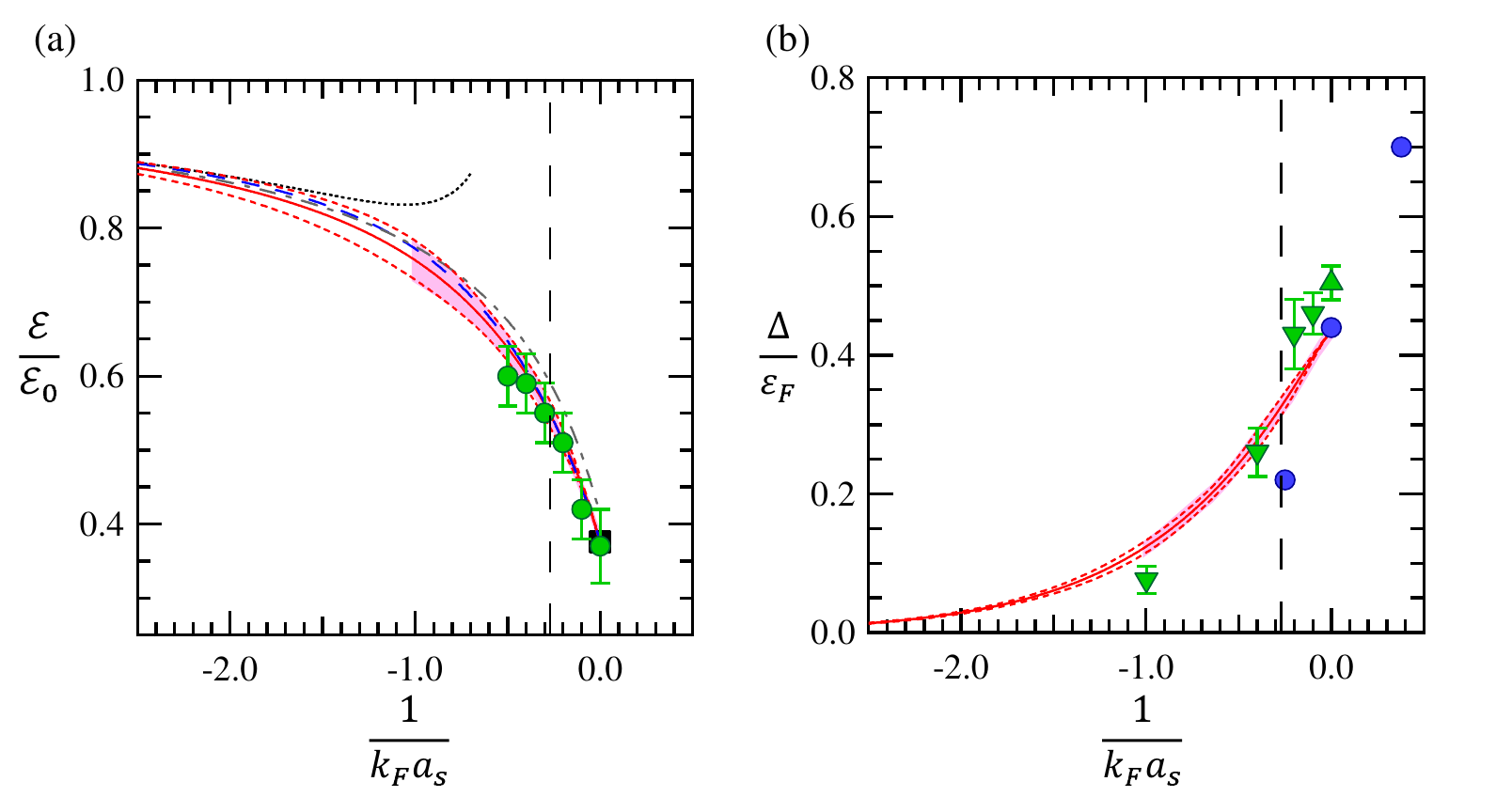}}
\caption{(color online)
Universal dimensionless EOS of spin-1/2 Fermi system in the ground state interacting with an s-wave scattering length at the zero range limit ($k_Fr_e=0$).
The horizontal axis shows the interaction parameter $1/k_Fa_s$ and the vertical long-dashed lines indicate the characteristic interaction parameter $(1/k_Fa_s)_c=-0.27$ for neutron matter, which is shown in Fig. \ref{Fig2}(b).
The pink bands show experimental values, and the width of the bands shows the range of experimental error.
The red curves and the red dashed curves are the approximated EOS and the error ranges from the Pad{\'e} approximation given in Eq. (\ref{eq46}).
(a) Universal EOS for the dimensionless internal energy density.
The black square at the unitarity limit \cite{56} and the gray dash-dotted curve \cite{60} are experimental data measured by different methods.
The green circles are values calculated by a QMC method \cite{61}.
The blue dashed curve is a theoretical EOS given by ETMA \cite{62}.
The black dotted curve is the asymptotic EOS given by Eq. (\ref{eq27}) up to the second term.
(b) Universal EOS for the dimensionless superfluid gap.
The blue circles are experimental data measured by a spectroscopic method \cite{63}.
The green triangles are values calculated by a QMC method \cite{64, 65}.
}
\label{Fig7}
\end{figure}

All of the experimental data and their validity are presented in Ref.~\refcite{53}.
In this review paper, we consider the EOS of the internal energy density and the superfluid gap, which are especially relevant for neutron matter.
The experimentally determined EOS for the internal energy density in the superfluid state is shown in Fig. \ref{Fig7}(a) as the pink band.
The width of the band shows the range of the experimental error.
The temperature parameter of the Fermi system is $T/T_F=0.06$ in this experiment.
To satisfy the condition of the zero-temperature limit, the fermions should be in the superfluid state, which has zero entropy.
Since the superfluid transition temperature decreases as the interaction parameter decreases \cite{59}, the Fermi system cannot realize the superfluid state below a certain level of the interaction parameters.
Under our experimental conditions, the Fermi system realizes the superfluid state in the region $1/k_Fa_s>-1.14$.
Our data is consistent with experimental data measured at the unitarity limit \cite{56}, as shown by the black square. It is also qualitatively consistent with an EOS measured by using a different method \cite{60}, as shown by the gray dashed-dotted line.
The green circles and the blue dashed curve are theoretical calculations by the quantum Monte-Carlo method \cite{61} and the extended T-matrix approximation \cite{62}, respectively.
They show good agreement with our data.
The black dotted curve shows the asymptotic EOS at the BCS limit, which is calculated up to the third term of Eq. (\ref{eq27}).
The vertical long-dashed lines indicate the characteristic interaction parameter $(1/k_Fa_s)_c=-0.27$ for neutron matter, which is shown in Fig. \ref{Fig2}(b).

Since we want a continuous EOS from the BCS limit to the unitarity limit, we interpolate the experimental data around the unitarity limit to the asymptotic EOS at the BCS limit using the Pad{\'e} approximation. 
We fit our experimental data to the function
\begin{equation}
f_\mathcal{E}^{\rm{Pade}}(\tilde{x})=\frac{\tilde{x}^3+\alpha_1\tilde{x}^2+\alpha_2\tilde{x}+\alpha_3}{\tilde{x}^3+\alpha_4\tilde{x}^2+\alpha_5\tilde{x}+\alpha_6},
\label{eq46}
\end{equation}
with $\tilde{x}=\frac{1}{k_Fa_s}$.
This function can be expanded at the BCS limit as
\begin{equation}
f_\mathcal{E}^{\rm{BCS}}(\tilde{x}\rightarrow-\infty)=1+(\alpha_1-\alpha_4)\tilde{x}^{-1}+(\alpha_2-\alpha_1\alpha_4+\alpha_4^2-\alpha_5)\tilde{x}^{-2}+\dots,
\label{eq47}
\end{equation}
and it can be expanded at the unitarity limit as
\begin{equation}
f_\mathcal{E}^{\rm{Pade}}(\tilde{x}\rightarrow0)=\frac{\alpha_3}{\alpha_6}+\frac{\alpha_2\alpha_6-\alpha_3\alpha_5}{\alpha_6^2}\tilde{x}+\dots.
\label{eq48}
\end{equation}
By a comparison to Eqs. (\ref{eq26}) and (\ref{eq27}) up to the third term, we can constrain the coefficients of Eq. (\ref{eq46}) to $(\alpha_1-\alpha_4)=\frac{10}{9\pi}$, $(\alpha_2-\alpha_1\alpha_4+\alpha_4^2-\alpha_5)=\frac{4(11-2\rm{ln}2)}{21\pi ^2}$, $\frac{\alpha_3}{\alpha_6}=\xi$, and $\frac{\alpha_2\alpha_6-\alpha_3\alpha_5}{\alpha_6^2}=-\frac{5\pi}{2}\frac{\mathcal{C}}{k_F^4}$.
Since we can fix the values of $\xi$ and $\frac{\mathcal{C}}{k_F^4}$ to our experimental values of $\xi=0.375(10)$ and $\frac{\mathcal{C}}{k_F^4}=0.1141(15)$, the only fitting parameters are $\alpha_4$ and $\alpha_6$.
The approximated EOS given by the fitting is shown in the figure by the red curve.
The parameters for Eq. (\ref{eq46}) are $\alpha_1=7.81\times 10^{-3}$, $\alpha_2=0.517$, $\alpha_3=-0.145$, $\alpha_4=-0.346$, $\alpha_5=0.454$, and $\alpha_6=-0.387$.
The red dashed curves above and below the red curve show the error range of the fitting.
We consider the experimental errors in the fitting procedure except for the error in $\xi$, because the fitting becomes unstable.
Since the relative error in $\xi$ is 2.7\%, there is an additional relative error in the approximated curve.

Let the dimensionless chemical potential be $f_{\mu}=\mu / \varepsilon_{\rm F}(n)$.
Then, the EOS is given by
\begin{equation}
f_\mu(\tilde{x})=f_\mathcal{E}(\tilde{x})-\frac{\tilde{x}}{5}\frac{df_\mathcal{E}(\tilde{x})}{d\tilde{x}}.
\label{eq49}
\end{equation}
By defining the dimensionless superfluid gap as  $f_{\Delta}=\Delta / \varepsilon_{\rm F}(n)$, we can estimate the EOS from $f_\mu(\tilde{x})$ by solving the following standard gap equation:
\begin{equation}
1=-\frac{2}{\pi}\frac{\sqrt{f_{\Delta}}}{\tilde{x}}I_1\left( \frac{f_{\mu}(\tilde{x})}{f_{\Delta}} \right),
\label{eq50}
\end{equation}
where the function of $I_1$ is defined in Ref.~\refcite{62.5}.
Fig. \ref{Fig7}(b) shows the EOS for the dimensionless superfluid gap given by the experimental data and the Pad{\'e} approximation, by the pink band and the red curve, respectively.
The width of the band and the red dashed curves show the error ranges.
The blue circles are experimental values measured by a spectroscopic method \cite{63}.
The green triangles are values calculated by a quantum Monte Carlo method \cite{64,65}.

We show two EOSes under the condition of the zero range limit ($k_Fr_e=0$) in Fig. \ref{Fig7}.
When the effective range has a non-negligible positive value ($k_Fr_e>0$), the effective range prevents the formation of cooper pairs because the attractive interaction is weakened by the positive effective range, as discussed in Sec. \ref{Sec2}.
This results in an increase in the internal energy by a decrease in the gap energy.
On the other hand, it has been pointed out that the effective range also decreases the internal energy by increasing the Hartree energy \cite{66}.
Since the net change of the internal energy depends on the value of the interaction parameter, the EOS shown in Fig. \ref{Fig7}(a) cannot constrain the EOS for the internal energy density of neutron matter.
However, the EOS shown in Fig. \ref{Fig7}(b) gives an upper bound on the EOS for the s-wave superfluid gap realized in neutron matter.

\begin{figure}[th]
\centerline{\includegraphics[width=12.5cm]{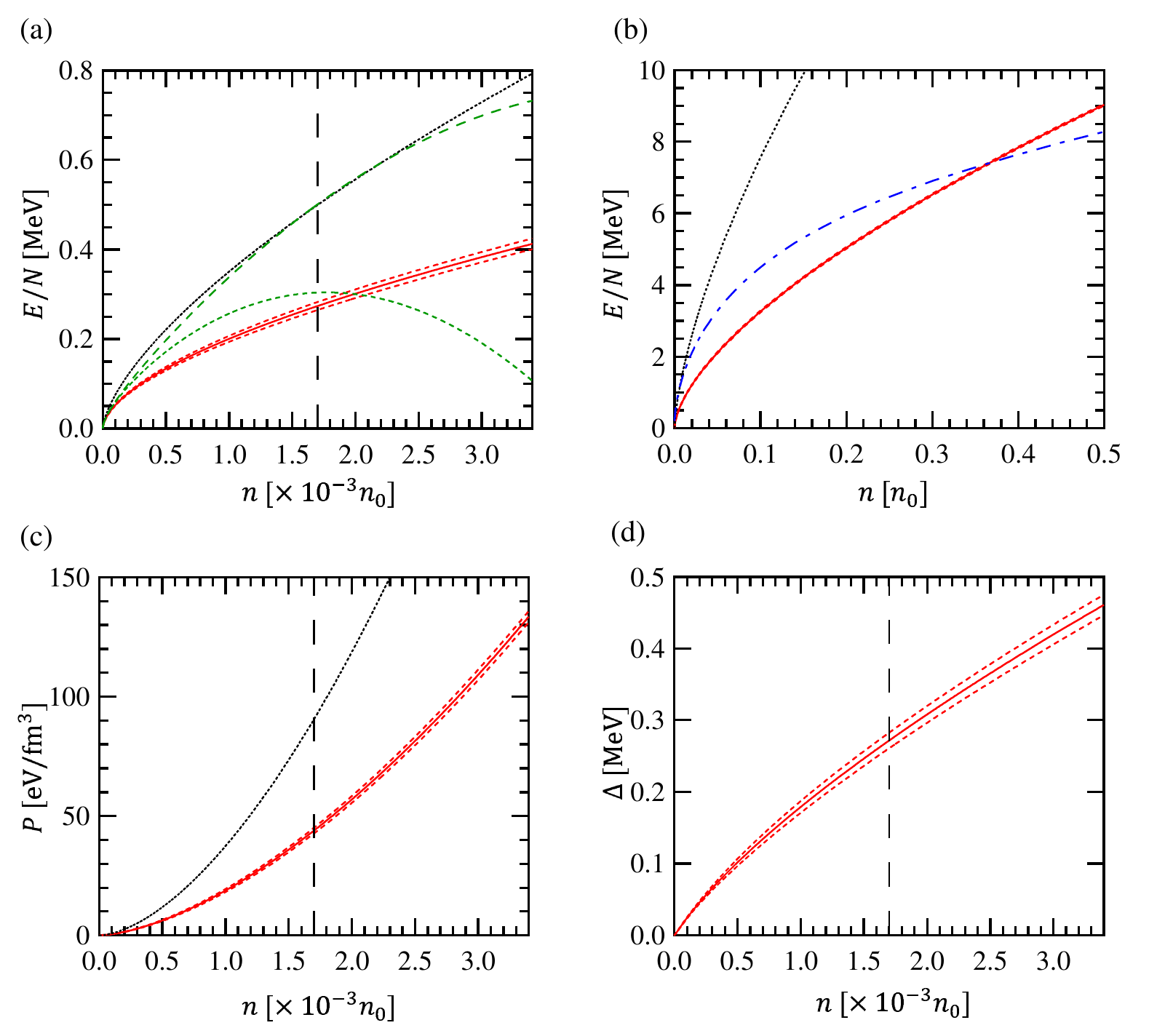}}
\caption{(color online)
EOS of dilute neutron matter determined by the cold-atom quantum simulator at the zero-range limit ($k_Fr_e=0$).
The horizontal axes show the neutron number density in units of the nuclear saturation density $n_0$.
The vertical long-dashed lines indicate the characteristic density at $n=n_c$.
The red curves with the dashed red curves show the EOS and the error range, taken from the EOSes shown in Fig. (\ref{Fig7}).
The black curves show the EOS of non-interacting spin-1/2 fermions.
(a) Internal energy per neutron in the density region $0<n<2n_c$.
The green short-dashed curve and the green long-dashed curve are asymptotic curves given by Eq. (\ref{eq26}) with $r_e=0$ and with $r_e=2.75$~fm, respectively.
(b) Internal energy per neutron in the density region $0<n<0.5n_0$.
The blue dashed curve is a theoretical calculation in the theoretical model of AV18 \cite{67}.
(c) Pressure in the density region $0<n<2n_c$.
(d) Superfluid gap in the density region $0<n<2n_c$.
}
\label{Fig8}
\end{figure}

The EOS of dilute neutron matter can be directly converted from the EOSs shown in Fig. \ref{Fig7}.
We apply the determined EOS $f_\mathcal{E}^{\rm{Pade}}$, the mass of a neutron, the neutron--neutron s-wave scattering length to Eq. (\ref{eq28}), and calculate the internal energy per neutron as a function of the neutron density, under the condition $k_Fr_e=0$.
We show the EOS in the density region $n<2n_c$ in Fig. \ref{Fig8}(a).
The horizontal axes are the neutron density in units of 0.1~\% of the nuclear saturation density $n_0$.
The black dotted curves show the EOS of the non-interacting spin-1/2 fermions.
It is interesting that the internal energy decreases dramatically just by working the s-wave scattering length of $a_s=-18.5$~fm to fermions.
The vertical black dashed line indicates the characteristic density at $n=n_c$.
The determined EOS in the range $n<n_c$ gives the quantitative behavior of the exact EOS for neutron matter, where the influence of the effective range is not significant.
We also show the asymptotic EOS at the low density limit, which is given by Eq. (\ref{eq26}).
The green short-dashed curve is the asymptotic EOS calculated with $r_e=0$, and the green long-dashed curve is the asymptotic EOS calculated with $r_e=2.75$~fm.
The asymptotic EOS with the positive effective range shows a higher energy than that with zero effective range.
At a density around $n\sim n_c$, the value of $|k_Fa_s|=3.7$ is already larger than 1,
and hence the asymptotic EOS is not appropriate at this density.
We show the determined EOS in the density region $n<2n_0$ in Fig. \ref{Fig8}(b), where the influence of the effective range is dominant.
For comparison, we include a theoretical curve of the model of AV18 \cite{67} shown by the blue dashed curve with the zero-range EOS.
While the two curves look similar, the physics are different.
In the case of the EOS with zero-range condition, the Fermi system approaches the unitarity limit and the fermions have a large superfluid gap.
On the other hand, in the case of the EOS with the positive effective range, the fermions have a smaller superfluid gap.
In Fig. \ref{Fig8}(c) and (d), we show the pressure and superfluid gap of neutron matter at low density.
The pressure is calculated from the internal energy shown in \ref{Fig8}(a) using Eq. (\ref{eq29}).
The superfluid gap is calculated from the EOS $f_\Delta(1/k_Fa_s)$ shown in \ref{Fig7}(b) by $\Delta(n)=\varepsilon_{\rm F}(n)f_\Delta\left(\frac{1}{k_F(n)a_s}\right)$.
This curve determined under the condition of zero-range gives the upper bound on the real s-wave superfluid gap of neutron matter with a positive finite range.

\section{Summary \label{Sec6}}

We introduced a cold-atom quantum simulation for Fermi systems interacting via s-wave scattering at the zero-temperature limit to better understand the fundamental physics of dilute neutron matter.
When the interaction potential between particles has a short-range shape and the wavelength of the matter wave, which is given by the temperature and the density, is longer than the interaction range of the potential, the details of the  interaction potential is reduced to two parameters: the s-wave scattering length and the effective range.
Based on this, we can formulate completely different particle systems that show common universal physical phenomena that do not depend on the details of the particles and the absolute length and the energy scale.
We demonstrated that the two parameters can be included in a grand canonical Hamiltonian using the concept of a pseudo-potential.
We also found that the two parameters work as thermodynamic variables according to the Hellmann--Feynman theorem.
By considering dimensionless thermodynamic variables, we showed that the important physical quantities are relative values that can be normalized by a reference scale of the system.
In this experiment, the EOS of a spin-1/2 Fermi system was determined in the interaction region $1/k_Fa_s \leq 0$ with $k_Fr_e=0$ at the zero-temperature limit.
The EOS gives the fundamental EOS of neutron matter at a dilute density up to 0.2\% of the nuclear saturation density, where the influence of the effective range is not significant.

Recently, a method of optical control of the Feshbach resonance has been developed \cite{68}.
This novel technique has potential for controlling the s-wave scattering length and the effective range independently.
By using this method, we can follow the red dotted arrow in Fig. \ref{Fig1} and extend the EOS to higher densities up to $n_0/2$, where the neutron--neutron interaction can be basically described by the two parameters.
At higher densities than $n_0/2$, the higher partial waves contribute to the EOS \cite{69}.
Therefore, a quantum simulation using the p-wave Feshbach resonance is also important \cite{70,71}.
The EOS can also be extended to a higher density by a theoretical method.
The proper theoretical frameworks for the quantum many-body system were extracted in this work.
By tuning the Hamiltonian to reproduce the energy dependent collisional phase shift, it is possible to calculate the EOS including the higher partial waves.
Also, it is possible to include protons by considering a four-component Fermi system.
In this way, this experimental data is valuable for testing many-body theories.

\section*{Acknowledgements}

The present study was supported by Grants-in-Aid for Scientific Research on Innovative Areas No. 24105001, No. 24105006, and No. 18H05406.

\end{document}